\documentclass[12pt]{article}
\usepackage[a4paper]{geometry}
\usepackage[italian,english]{babel}

\usepackage{fancyhdr}
\usepackage{amsmath, accents}
\usepackage{amsfonts}
\usepackage{amstext}
\usepackage{amssymb}
\usepackage{amsthm}
\usepackage{amscd}
\usepackage{mathrsfs}

\usepackage[pagebackref,draft=false]{hyperref}
\hypersetup{colorlinks,
linkcolor=myrefcolor,
citecolor=mycitecolor,
urlcolor=myurlcolor}

\usepackage[capitalize]{cleveref}
\usepackage{caption}
\usepackage{etaremune}

\usepackage{xcolor}
\definecolor{myurlcolor}{rgb}{0,0,0.4}
\definecolor{mycitecolor}{rgb}{0,0.5,0}
\definecolor{myrefcolor}{rgb}{0.5,0,0}
\usepackage{graphicx}
\usepackage{tikz}
\usepackage{tikz-cd}

\usepackage{etoolbox}
\usepackage{makeidx}
\usepackage{sectsty}
\usepackage{dsfont}
\usepackage{enumitem} 
\usepackage[]{latexsym}
\usepackage{braket}
\usepackage{caption}
\usepackage[utf8]{inputenx}
\usepackage[T1]{fontenc}
\usepackage{lmodern}
\usepackage{textcomp}
\usepackage{microtype}
\usepackage{totcount}
\usepackage{blindtext}

\newtheorem{remark}{Remark}

\newtheorem*{proof*}{Proof}

\newcommand{\be}{\begin{equation}}
\newcommand{\ee}{\end{equation}}
\newcommand{\bea}{\begin{eqnarray}}
\newcommand{\eea}{\end{eqnarray}}
\newcommand{\vsp}{\vspace{0.4cm}}

\newcommand{\blue}[1]{\textcolor{blue}{{#1}}}

\newcommand{\lra}{\longrightarrow}

\newcommand{\hh}{\mathcal{H}}
\newcommand{\bh}{\mathcal{B}(\mathcal{H})}

\newcommand{\stsp}{\mathscr{S}}

\newcommand{\X}{\mathbb{X}}
\newcommand{\Y}{\mathbb{Y}}
\newcommand{\p}{\mathbb{P}}
\newcommand{\Q}{\mathbb{Q}}

\newcommand{\dd}{{\rm d}}

\newcommand{\nn}{\nonumber}

\title{Geometry from divergence functions and complex structures}

\author{F. M. Ciaglia$^{1,6}$  \href{https://orcid.org/0000-0002-8987-1181}{\includegraphics[scale=0.7]{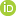}}, F. Di Cosmo$^{2,3,7}$ \href{https://orcid.org/0000-0002-8987-1181}{\includegraphics[scale=0.7]{ORCID.png}}, A. Figueroa$^{4,8}$, \\ G. Marmo$^{4,5,9}$ \href{https://orcid.org/0000-0003-2662-2193}{\includegraphics[scale=0.7]{ORCID.png}}, L. Schiavone$^{4,5,10}$ \href{https://orcid.org/0000-0002-1817-5752}{\includegraphics[scale=0.7]{ORCID.png}}, \\ 
\footnotesize{$^{1}$\textit{ Max Planck Institute for Mathematics in the Sciences, Leipzig, Germany}} \\
\footnotesize{$^{2}$\textit{ ICMAT, Instituto de Ciencias Matem\'{a}ticas (CSIC-UAM-UC3M-UCM)}} \\
\footnotesize{$^{3}$\textit{ Depto. de Matem\'aticas, Univ. Carlos III de Madrid, Legan\'es, Madrid, Spain}}  \\
\footnotesize{$^{4}$\textit{ Dipartimento di Fisica ``E. Pancini'', Universit\`a di Napoli Federico II,  Naples, Italy}} \\
\footnotesize{$^{5}$\textit{ INFN-Sezione di Napoli, Naples, Italy}} \\
\footnotesize{$^{6}$\textit{ e-mail: \texttt{florio.m.ciaglia[at]gmail.com} and \texttt{ciaglia[at]mis.mpg.de}}} \\
\footnotesize{$^{7}$\textit{ e-mail: \texttt{fabiodicosmo[at]gmail.com}}} \\
\footnotesize{$^{8}$\textit{ e-mail: \texttt{figueroarmando[at]yahoo.com.mx}}} \\
\footnotesize{$^{9}$\textit{ e-mail: \texttt{marmo[at]na.infn.it}}} \\
\footnotesize{$^{10}$\textit{ e-mail: \texttt{lucaschiavone[at]live.it}}} 
}

\date{}

\begin{document}

\maketitle

\begin{abstract}
Motivated by the geometrical structures of quantum mechanics, we introduce an almost-complex structure $J$ on the product $M\times M$ of any parallelizable statistical manifold $M$. 
Then, we use $J$ to extract a pre-symplectic form and a metric-like tensor on $M\times M$  from a divergence function.
These tensors may be pulled back to $M$, and we compute them in the case of an N-dimensional symplex with respect to the Kullback-Leibler relative entropy, and in the case of (a suitable unfolding space of) the manifold of faithful density operators with respect to the von Neumann-Umegaki relative entropy.
\end{abstract}

\tableofcontents

\thispagestyle{fancy}
\section{Introduction}

In information geometry, it is customary to consider Riemannian metric tensors on (suitable submanifolds of) the space of probability distributions on some measure space in order to introduce a notion of distance or distinguishability among different probability distributions. 
The idea of distance between probability distributions goes back to Fisher and has been elaborated by Rao \cite{Rao-1945},   Cencov \cite{Cencov-1982},  and Amari and Nagaoka \cite{A-N-2000}, to name just a few.

Let us briefly recall the classical setting in the simple case of a discrete, finite sample space $\mathcal{X}=\{1,...N\}$.
In this case, an arbitrary probability distribution  on $\mathcal{X}$ may be identified with a probability vector $\mathbf{p}=(p_{1},...,p_{N})$, where   $ p_{j}\in[0,1]$ for all $j=1,...,N$, and $\sum_j p_j = 1 $. 
The collection of probability distributions is thus in one-to-one correspondence with an $(N-1)$-dimensional simplex $\Delta$ sitting in $\mathbb{R}^{N}$.
The open interior $\Delta_{+}$ of $\Delta$ made up of all those probability vectors with strictly positive components is a smooth manifold of dimension $(N-1)$, and  the Fisher-Rao metric tensor on it has the form
\begin{eqnarray}\label{eqn: F-R metric on the simplex}
g_{FR} = \sum_j p^j d \ln p^j \otimes d \ln p^j \,=\,\sum_{j}\,\frac{1}{p^{j}}\,\mathrm{d}p^{j}\otimes\mathrm{d}p^{j} ,
\end{eqnarray}
which is related to (four times) the round metric tensor on (an open submanifold of) the $n$-dimensional sphere, $ g = 4\sum_j dx_j \otimes dx_j$, where $ x_j = \sqrt{p_j} $.

In the case of a discrete, finite measure space, an important theorem by Cencov states that if we ask the distinguishability between probability distributions  not to increase under the action of stochastic maps,  then the metric tensor is necessarily a multiple of the so-called Fisher-Rao tensor.
This theorem has been generalized also to the case of a non-discrete measure space  provided some additional conditions are met \cite{A-J-L-S-2017,B-B-M-2016}.

In the quantum case, the situation is completely different.
Indeed, when we pass from probability distributions to quantum states, that is, to density operators on the Hilbert space of the system, already in the finite-dimensional case, it is possible to prove that Cencov's theorem is maximally violated in the sense that there is an infinite number of metric tensors satisfying the quantum analogue of the monotonicity property under classical stochastic maps \cite{C-M-1991,Petz-1996}.
This means that, in the quantum case, there is additional freedom  in choosing a relevant metric tensor as long as the classical Fisher-Rao metric tensor is recovered when we perform a quantum-to-classical limit.
For instance, as we will recall in section \ref{sec: pure states} and section \ref{sec: mixed states}, this is precisely what happens for the Fubini-Study metric tensor on the space of pure quantum states, and with the metric tensor on the space of (faithful) density operators which is associated with the von Neumann-Umegaki relative entropy.

In this contribution, we will review the geometrical aspects of classical and quantum information theory, and we will exploit the parallel between classical and quantum information geometry to argue that the geometry of the quantum case leads to the definition of additional geometric structures in the classical case.
Specifically, we will take inspiration from the  geometry of the pure quantum states in order to build an almost complex structure on the product $M\times M$ of any parallelizable statistical manifold $M$ by means of which we may extract a pre-symplectic form and a symmetric $(0,2)$ tensor field on $M\times M$ starting from a divergence function (relative entropy).
In particular, we will compute the pre-symplectic form on the $N$-simplex which is associated with the Kullback-Leibler relative entropy.
Furthermore, we will consider also the case of faithful density operators, and compute the metric tensor on $M$ associated with the von Neumann-Umegaki relative entropy.

\section{Remarks on the geometry of pure states}\label{sec: pure states}

Here, we will recall some of the basic ingredients of the so-called geometrization of quantum mechanics \cite{A-S-1999,C-CG-M-2007,Kibble-1979}.
We will focus on two aspects which will be further explored in the following sections.

On the one hand, we will introduce the idea according to which it is possible to recover the geometrical structures of the classical case, e.g., the Fisher-Rao metric tensor, starting from the quantum ones by means of a suitable immersion of probability distributions into quantum states.
In this section, we will develop this idea in the context of pure states, while in section \ref{sec: mixed states}, we will present an extension of this idea in the context of mixed states.

On the other hand, we will exploit the geometrical structure of the space of pure quantum states in order to highlight the role of the complex structure in the definition of geometrical tensors starting from functions.
In section \ref{sec: statistical manifolds}, we will start from this idea in order to reformulate what is usually done in the context of information geometry by introducing an almost-complex  structure on the Cartesian product $M\times M$ of a statistical manifold $M$ with itself.
In this way, we will be able to define metric-like and a pre-symplectic tensor on $M\times M$ that may be pulled back to $M$.

\vsp

In standard quantum mechanics \cite{Dirac-1964,E-M-M-S-2014}, a Hilbert space $\hh$ is associated with a quantum system, and observables are identified with self-adjoint operators on $\hh$.
According to Dirac, the linear structure of $\hh$ is crucial to describe the superposition principle of quantum mechanics.
Specifically, if $\psi,\phi\in\hh$ are vectors representing two states of the system, the linear structure of $\hh$ allows to say that $\psi+\phi\in\hh$ represents another admissible state for the system.
This way of looking at states as vectors in $\hh$ may be satisfying from the point of view of the superposition principle, but it is not fully compatible with the statistical content of quantum mechanics.
Indeed, one of the fundamental prescription in quantum mechanics is that the quantity
\be\label{eqn: expectation value of A on psi}
\langle A \rangle_{\psi}\,:=\,\langle\psi |A| \psi\rangle,
\ee
where $A$ is a self-adjoint linear operator on $\hh$ and $\langle,\rangle$ is the Hilbert product on $\hh$, has to be interpreted as the expectation value of the observable $A$ on the state $\psi$.
Then, relying on the spectral decomposition of $A$ given by 
\be
A\,=\,\int_{\sigma(A)}\,\lambda\,\mathrm{d}E_{A}(\lambda),
\ee
where $\sigma(A)\subseteq\mathbb{R}$ is the spectrum of $A$ and $E_{A}$ is the projection-valued measure associated with $A$ \cite{R-S-1980-1}, the quantity 
\be
\mu(\psi , E_{A}(O))\,=\,\langle \psi |E_{A}(O)|\psi\rangle,
\label{eqn: probability measure associated with a self-adjoint operator}
\ee
where $O$ is a measurable subset of $\sigma(A)$, is interpreted as the probability that a measure of  $A$ on  $\psi$ gives an outcome which is in $O$.
Therefore, in order to make this picture consistent, we must have that 
\be
\mu(\psi , E_{A}(\sigma(A)))\,=\,1,
\ee
and thus, since $E(\sigma(A))$ is the identity operator $\mathbb{I}$ on $\hh$ for every observable $A$, we must have that
\be
\langle \psi,\psi\rangle\,=\,1,
\ee
that is, the vector $\psi$ must be normalized.
Consequently, the probabilistic-statistical interpretation of quantum mechanics forces us to leave the linear space $\hh$ and pass to the nonlinear manifold given by normalized vectors in $\hh$, that is, on the unit-sphere in $\hh$.
However, this is not the end of the story.
Indeed, if we look at equation \eqref{eqn: probability measure associated with a self-adjoint operator}, we can notice that the measure $\mu(\psi , E_{A}(O))$ is equal to the measure $\mu(\phi , E_{A}(O))$ if we consider the vector $\phi:=\mathrm{e}^{\imath\theta}\psi$ with $\theta\in\mathbb{R}$, and that $\phi$ is still a normalized vector in $\hh$.
This means that we have an additional $U(1)$ symmetry of which we can dispose of, so that the mathematical object that correctly describes a (pure) quantum state is the equivalence class $[\psi]$ associated with $\psi$ with respect to the action of $C_{0}=\mathbb{R}^{+}\times U(1)$ on $\hh$ given by scalar multiplication.
Eventually, we obtain that the statistical interpretation of quantum mechanics forces us to describe (pure) quantum states as points in the complex projective space $\mathbb{P}(\hh)\,=\,\hh_{0}/\mathbb{C}_{0}$ associated with $\hh$.

On this nonlinear manifold, the superposition principle is clearly not applicable in the same way as it is on $\hh$, however, it can be proved \cite{M-M-S-Z-2000} that there is a formulation of the superposition principle on $\mathbb{P}(\hh)$ which requires the specification of a third reference state.
From this point of view, the Hilbert space $\hh$ seems to be a very useful computational tool to express the superposition principle, and this simplicity is gained at the expenses of a redundant description of quantum states.

Once we accept that (pure) states in quantum mechanics are points in $\mathbb{P}(\hh)$, we may start to uncover the geometrical structures that are ``naturally'' present on this manifold.
To avoid technical difficulties, in the sequel we shall always assume $ \mathrm{dim}(\hh) = N <\infty$.

On the one hand, the Hermitian product of $\hh$ does not play any role in defining the manifold of pure states, it is only the action of $\mathbb{C}_{0}$ on the vector space $\mathbb{V}$ underlying $\hh$ that enters the game.
From the mathematical point of view, $\hh$ is a $N$-dimensional, complex vector space, say $\mathbb{V}$, endowed with an Hermitian product denoted by $\langle \psi, \phi\rangle$ which is, by convention, $\mathbb{C}$-linear with respect to the second entry and anti-linear with respect to the first entry. 
The group of isometries of $\langle , \rangle$ is a compact subgroup of the complex, general linear group of $\mathbb{V}$ called the unitary group and denoted by $\mathcal{U}(\hh)$ in order to emphasize that its definition depends on the Hermitian structure $\langle , \rangle$ on $\mathbb{V}$ (contrarily to the complex, general linear group which is defined for a generic complex vector space $\mathbb{V}$).
 
If $\{e_j\}_{j=1,\dots,N}$ is an Hermitian basis for $\hh=(\mathbb{V}, \langle~,~\rangle )$, the corresponding coordinates for an element $\psi$ are written as $\langle e_j,\psi\rangle\,=\,q_j+ip_j$ with $(q_j,p_j)$ real numbers. 
Therefore, the Hilbert space $\hh$ can be studied as a real, $2N$-dimensional linear manifold with a global coordinate chart given as above.
The smooth action of $\mathbb{C}_{0}$ on $\mathbb{V}$ is given by $\psi\mapsto \alpha\,\psi$ with $\alpha\in\mathbb{C}_{0}$, and the infinitesimal generators of the action are the linear vector fields
\be
\begin{split}
\Delta&\,=\,q^{j}\,\frac{\partial}{\partial q^{j}} + p^{j}\,\frac{\partial}{\partial p^{j}} \\
\Gamma&\,=\,p^{j}\,\frac{\partial}{\partial q^{j}} - q^{j}\,\frac{\partial}{\partial p^{j}} .
\end{split}
\ee
The  vector field $\Gamma$ implements the phase rotations, while $\Delta$ implements the dilations and describes the linear structure of the underlying vector space $\mathbb{V}$ \cite{C-I-M-M-2015}.
The vector fields $\Delta$ and $\Gamma$ determine  an involutive distribution $\mathcal{D}$ (they commute because $\mathbb{C}_{0}$ is Abelian), and they are complete because they are linear vector fields.
However, we need to discard the null vector of $\mathbb{V}$ (the unique fixed point of $\Delta$ and $\Gamma$)  in order for the quotient with respect to the action of $\mathbb{C}_{0}$ to be a smooth manifold. In the following $\mathbb{V}_0$ will denote the space obtained from the vector space $\mathbb{V}$ after removing the null vector. 
The resulting space $\mathbb{V}_{0}/\mathbb{C}_{0}$ is the so-called complex projective space for the complex vector space $\mathbb{V}$, and we denote it by $\mathbb{CP}(\mathbb{V})$ in order to emphasize the fact that the manifold structure of the complex projective space depends on the complex vector space structure of $\mathbb{V}$ and does not depend on the Hermitian product turning $\mathbb{V}$ into the Hilbert space $\hh$.
The canonical projection map from $\mathbb{V}$ to $\mathbb{CP}(\mathbb{V})$ will be denoted by $\pi$.

On the other hand, the Hermitian product $\langle , \rangle$ determines a Hermitian tensor $H$ that reads
\be
H\,=\,\sum_{j=1}^{N}\left(\dd q_j\otimes\dd q_j\,+\,\dd p_j\otimes\dd p_j\right)\,+\,i\,\left(\dd q_j\otimes\dd p_j\,-\,\dd p_j\otimes\dd q_j\right)\,=\,\rm g\,+\,i\,\omega ,
\ee
where $\rm g$ is a Riemannian metric, and $\omega$ a symplectic structure.
The $(1,1)$ tensor field
\begin{equation}\label{eqn: complex structure on H0}
J=\dd q^{k}\otimes\frac{\partial }{\partial p^{k}} - \dd p^{k}\otimes\frac{\partial}{\partial q^{k}}
\end{equation}%
is such that $J^{2}=-\mathrm{Id}$, and it determines a complex structure compatible with $\rm g$ and $\omega$ in the sense that 
\begin{align}
&\mathrm g(Ju,v)\,=\,\omega(u,v), \nn \\ &\mathrm g(Ju,Jv)\,=\,\mathrm g(u,v),  \nn \\ & \omega(Ju,Jv)\,=\,\omega(u,v)
\label{eq4}
\end{align}
for every couple $(v, u)$ of vector fields.
The triple $(J, \rm g, \omega)$ determines a K\"{a}hler structure on $\mathbb{V}$.
Upon considering the contravariant tensor fields $\Lambda=\omega^{-1}$, $\rm G =\rm g^{-1}$, it is possible to show  \cite{M-Z-2018} that $\widetilde{\Lambda}=R\Lambda$, $\widetilde{\rm G}=R\,\rm G$  with $R=\rm g(\Delta,\Delta)$, are tensor fields ``projectable'' with respect to the projection map $\pi$.
This means that there are two tensor fields $\Lambda_{\pi}$, and $\rm G_{\pi}$ on $\mathbb{CP}(\mathbb{V})$ such that $\widetilde{\Lambda}$ is $\pi$-related with $\Lambda_{\pi}$, $\widetilde{\rm G}$ is $\pi$-related with $\rm G_{\pi}$, and $J$ is $\pi$-related with $J_{\pi}$ \cite{M-Z-2018}.
Furthermore, $\Lambda_{\pi}$ and $\rm G_{\pi}$ are invertible their inverses are a simplectic form $\omega_{\pi}$ and a Riemannian metric tensor $g_{\pi}$ on $\mathbb{CP}(\mathbb{V})$ (the so-called Fubini-Study metric on the complex projective space), respectively, and there is a complex structure $J_{\pi}$ on $\mathbb{CP}(\mathbb{V})$ that is compatible with $\omega_{\pi}$ and $g_{\pi}$  (see equation \eqref{eq4}).
Essentially, the triple $(J_{\pi}, \rm g_{\pi}, \omega_{\pi})$ determines a K\"{a}hler structure on $\mathbb{CP}(\mathbb{V})$ which clearly depends on the Hermitian product $\langle ,\rangle$.

Note that all the linear vector fields generating the smooth, left action of $\mathcal{U}(\hh)$ on $\mathbb{V}$ ($\psi\mapsto U\psi$) commute with $\Delta$ and $\Gamma$, and thus are ``projectable'' on $\mathbb{CP}(\mathbb{V})$ and determine a smooth left action of $\mathcal{U}(\hh)$ on $\mathbb{CP}(\mathbb{V})$.
It can be proved \cite{E-M-M-2010} that the Fubini-Study metric $g_{\pi}$ is the unique Riemannian metric on $\mathbb{CP}(\mathbb{V})$ which is invariant with respect to the action of $\mathcal{U}(\hh)$ up to a constant factor.

The space of pure quantum states is then the complex projective space $\mathbb{CP}(\mathbb{V})$ endowed with the K\"{a}hler structure given above, and will be denoted as $\mathbb{P}(\hh)$ to emphasize the fact that the K\"{a}hler structure depends on the Hilbert space $\hh$.

On the punctured Hilbert space $\hh_{0}=\hh-\{0\}$, we may define the following Hermitian tensor
\begin{eqnarray}\label{eqn: pullback of Hermitean tensor}
h  = \frac{\langle\mathrm{d}\psi|\mathrm{d}\psi\rangle}{\langle\psi|\psi\rangle} -\frac{\langle\mathrm{d}\psi|\psi\rangle\langle\psi|\mathrm{d}\psi\rangle}{\langle\psi|\psi\rangle^{2}}\, .
\end{eqnarray}
The relevance of this tensor stems from the fact that, quite interestingly, its real part  is the pullback to $\hh_{0}$ of the Fubini-Study metric $g_{\pi}$ on $\mathbb{P}(\hh)$, while its immaginary part is the pullback to $\hh_{0}$ of the symplectic form $\omega_{\pi}$ defining the canonical K\"{a}hler structure on $\mathbb{P}(\hh)$  \cite{E-M-M-2010}.
Consequently, we may look at $h $ as an unfolding tensor for $g_{\pi}$ and $\omega_{\pi}$, and this way of looking at $h $ is particularly relevant when we want to address the issue of recovering the classical case from the quantum one \cite{F-K-M-M-S-V-2010}.
Specifically, given an arbitrary probability vector $\mathbf{p}=(p_{1},...,p_{N})$, we consider a sort of complex-valued square root of $\mathbf{p}$  given by the complex vector $(e^{i \, \theta_{1} }\sqrt{p_{1}},..., e^{i \, \theta_{N} }\sqrt{p_{N}})$, that is, we replace  the probability vector  with a ``probability amplitude'' vector 
\be
\psi_{p} \,= \sum_{j=1}^{N}\,z_j \,|e_j\rangle\,=\,\sum_{j=1}^{N}\, e^{i \, \theta_{j} }\sqrt{p_{j}}\,|e_j\rangle,
\ee
From the mathematical point of view, we may interpret this procedure as a nonlinear change of coordinates in $\hh_{0}$ so that a direct computation reveals that, in this new coordinates system, the expression of $h $ is
\begin{eqnarray}
h  & = & \frac{1}{4} \left( \langle d \ln p \otimes d \ln p \rangle_p  - \langle d \ln p  \rangle_p \otimes \langle d \ln p  \rangle_p \right) + \langle d \theta \otimes d \theta \rangle_p - \langle d \theta \rangle_p \otimes \langle d \theta \rangle_p + \nonumber \\
 & + & \frac{i}{2} \left( \langle d \ln p \wedge d \theta\rangle_p  - \langle d \ln p \rangle_p \wedge \langle d \theta \rangle_p \right),
\end{eqnarray}
where $\langle \cdot \rangle_{p} $ denotes  the expectation value with respect to the probability distribution $ \mathbf{p} $.
From this, it follows that  the real part of $h $ is nothing but the Fisher-Rao metric tensor whenever $\mathrm{d}\theta=0$.
According to the results in \cite{F-K-M-M-S-V-2010}, this procedure may also be extended to the case where $\hh$ is an infinite-dimensional Hilbert space of square-integrable functions on some measure space.


\vsp

Another relevant aspect of the geometry of pure quantum states is related with the possibility of describing the real and immaginary part of the Hermitian tensor $h $ by means of a potential function and the complex structure $J$ on $\hh_{0}$ given in equation \eqref{eqn: complex structure on H0}.
Specifically, given any function $F$ on $\hh_{0}$, we define the $(0,2)$ tensor
\be
\omega_{F}\,:=\,\dd\,\dd_{J}F\,=\,\dd\,\left(\,J \,\circ\,\dd F\right)\,.
\ee
This covariant tensor is clearly closed, and is easily seen to be antisymmetric because it is the exterior differential of a 1-form.
Then, we may define a $(0,2)$ tensor field $g_{F}$ by setting
\be
g_{F}(X,Y)\,=\,\omega_{F}(X,J (Y))\,.
\ee
If  the function $F$ is such that 
\be\label{eqn: compatibility between F,J and omega}
\omega_{F}(X,J(Y))\,=\,-\omega_{F}(J(X),Y)\,,
\ee
then $g_{F}$ is a symmetric tensor, and the triple $(J,\omega_{F},g_{F})$ defines a sort of\footnote{As, in general, the two tensor fields may be degenerate, we need to ``enlarge'' the definition of triple defining the K\"{a}hler structure.} K\"{a}hler structure on $\hh_{0}$.
For instance,  we may consider the function
\be
F\,=\,\ln(\langle\psi|\psi\rangle),
\ee
and a direct computation shows that $\omega_{F}$ coincides with the imaginary part of $h $, while $g_{F}$ coincides with the real part of $h $ given in equation \eqref{eqn: pullback of Hermitean tensor}.

Note that $F$ is not the pullback of a function on the complex projective space $\mathbb{P}(\hh)$ because $\ln(\langle\psi|\psi\rangle)$ changes by an additional constant under dilation.
Furthermore, it should be clear that the procedure just outlined will work on any manifold $M$ admitting a $(1,1)$ tensor $J$ such that
\be
J^{2}\,=\,-\mathrm{Id}\,.
\ee
We will exploit this instance in the following section.

\section{Almost complex structures and statistical manifolds}\label{sec: statistical manifolds}

In the last part of the previous section, we saw how the complex structure $J$ allows to build a closed two-form and a symmetric $(0,2)$ tensor field on $\hh_{0}$  starting from a single real-valued, smooth function on the manifold.

Here, we will adapt this idea to the case of statistical manifolds  in the context of information geometry.
A (naked) statistical manifold $M$ is just a smooth manifold the points of which parametrize, in a one-to-one way, a subset of probability distributions on some given outcome space $\mathcal{X}$.
A typical example is given by the statistical manifold of Gaussian probability distributions on $\mathbb{R}$, where a point in $M$ is given by $(\mu,\sigma)$ where $\mu$ is the mean and $\sigma$ the variance of the Gaussian.

It is customary to fix a reference measure $\mu$ on $\mathcal{X}$ in such a way that the points in $M$ parametrize a family of probability distributions by means of the map
\be
m\,\mapsto\;p(x,m)\,\mathrm{d}\mu,
\ee
where $p(x,m)$ is a function on $\mathcal{X}$ depending parametrically on $m\in M$.
Then, this immersion of $M$ into the space of probability distributions on $\mathcal{X}$ allows to define a geometrical structure on $M$, namely, the Fisher-Rao metric tensor $g_{FR}$  given by
\be
(g_{FR})_{jk}(\xi(m))\,:=\,\int_{\mathcal{X}}\,p(x,\xi(m))\,\frac{\partial \ln (p(x,\xi(m)))}{\partial \xi^{j}}\,\frac{\partial \ln (p(x,\xi(m)))}{\partial \xi^{k}}\,\mathrm{d}\mu(x)\,,
\ee
where $\{\xi^{j}\}_{j=1,...,dim(M)}$ is a coordinate chart for $M$.
It can be proved that this expression transform as a $(0,2)$ tensor under coordinate change   \cite{A-J-L-S-2017}.

Quite remarkably, the Fisher-Rao metric tensor on $M$ may also be ``extracted'' from a suitable two-point function $D$ (i.e., a function on $M\times M$), called a divergence function, or a contrast function, which, in some cases, may be interpreted as a relative entropy.
Specifically, a divergence function $D$  is a real-valued, smooth function on $M\times M$ such that (see  section 3.2 in \cite{A-N-2000})
\be
D(m_{1},m_{2})\,\geq\,0\;\;\forall (m_{1},m_{2})\in M\times M,
\ee
and such that the equality in the previous equation holds iff $m_{1}=m_{2}$.

Starting from a divergence function $D$, it is possible to extract a metric tensor $g$ on $M$ by setting $g\,=\,g_{jk}\,\mathrm{d}q^{j}\otimes\mathrm{d}q^{k}$ with
\be
g_{jk}\,=\,\left.\left(\frac{\partial^{2} D}{\partial x^{j}\,\partial x^{k}}\right)\right|_{diag}\,=\, \left.\left(\frac{\partial^{2} D}{\partial y^{j}\,\partial y^{k}}\right)\right|_{diag}\,=\,-\left.\left(\frac{\partial^{2} D}{\partial x^{j}\,\partial y^{k}}\right)\right|_{diag},
\ee
where $(q^{j})$ is a coordinate chart on $M$,  $(x^{j},y^{j})$ is a coordinate chart on $M\times M$ which is adapted to the product structure of the manifold, and $\left.\right|_{diag}$ denotes the restriction to the diagonal of $M\times M$.
Note that these second derivatives patch together to form a $(0,2)$ tensor on $M$ because $D$ satisfies the properties characterizing a divergence function.

\vsp

A paradigmatic example of divergence function which is interpreted as a relative entropy is given by the Kullback-Leibler relative entropy
\be
S_{KL}(p,q)\,=\,\sum_{j=1}^{N}\,p^{j}\,\ln\left(\frac{p^{j}}{q^{j}}\right).
\ee
If we consider the statistical manifold $\Delta_{+}$, that is, the open interior of an $N$-dimensional simplex,  then a direct computation shows that the metric $g_{KL}$ we may extract from $S_{KL}$ is precisely the Fisher-Rao metric tensor given in equation \eqref{eqn: F-R metric on the simplex}.

\vsp

A coordinate-free formulation of this procedure is given in \cite{C-DC-L-M-M-V-V-2018,M-M-V-V-2017}, and a general formalism based on the Lie groupoid structure of $M\times M$ is presented in \cite{G-G-K-M-2019}.
Here, on the other hand, we would like to give an alternative (but equivalent), coordinate-free formulation of the extraction procedure which is inspired by K\"{a}hler geometry and thus will lead us to define also a pre-symplectic form on $M\times M$ starting from a divergence function.
Specifically, we assume that $M$ is a parallelizable manifold, so that we have a global basis $\{X_{j}\}$ of vector fields and its dually related global basis $\{\theta^{j}\}$ of differential 1-forms.
Note that for these bases to be dually related we must have that $\theta^{j}(X_{k}) = \delta^{j}_{k}$. 
Then, by denoting with $pr_{1}$ and $pr_{2}$ the left and right projections from $M\times M$ to $M$ respectively, we immediately obtain a global basis $\{\X_{j},\Y_{k}\}$ of vector fields  on $M\times M$ by setting
\be
\begin{split}
\X_{j}(pr_{1}^{*}(f))&\,=\,pr_{1}^{*}(X_{j}f)\;\;\;\forall \,f\in\mathcal{F}(M) \\
\Y_{j}(pr_{2}^{*}(f))&\,=\,pr_{2}^{*}(X_{j}f)\;\;\;\forall \,f\in\mathcal{F}(M).
\end{split}
\ee 
A global basis $\{\alpha^{j},\beta^{k}\}$ of 1-forms is obtained as the dually-related basis of $\{\X_{j},\Y_{k}\}$.
If  $i_{d}\colon M\lra M\times M$ is the diagonal immersion given by $m\,\mapsto\,i_{d}(m)=(m,m)$, it follows from proposition 2 in \cite{C-DC-L-M-M-V-V-2018} that $X_{j}$ is $i_{d}$-related with $\X_{j}+\Y_{j}$, and thus we have 
\be\label{eqn: important pullbacks}
i_{d}^{*}\alpha^{j}\,=\,i_{d}^{*}\beta^{j}\,=\,\theta^{j}.
\ee
If $(q^{r})$ is a coordinate chart on $M$ for which
\be
X_{j}\,=\,X_{j}^{r}\,\frac{\partial}{\partial q^{r}},\;\;\;\;\;\theta^{j}\,=\,\theta^{j}_{r}\,\mathrm{d}q^{r},
\ee
and $\{x^{r},y^{r}\}$ is a coordinate chart on $M\times M$ adapted to its product structure, we have
\be
\begin{split}
\X_{j}\,=\,X_{j}^{r}\,\frac{\partial}{\partial x^{r}},\;\;\;\;\;\alpha^{j}\,=\,\theta^{j}_{r}\,\mathrm{d}x^{r} \\
\Y_{j}\,=\,X_{j}^{r}\,\frac{\partial}{\partial y^{r}},\;\;\;\;\;\beta^{j}\,=\,\theta^{j}_{r}\,\mathrm{d}y^{r},
\end{split}
\ee
where, because duality, we must have $X^{r}_{j}\,\theta_{r}^{k}\,=\, \delta^{k}_{j}$.
Starting from these bases, it is possible to introduce an almost complex structure on $M\times M$, that is, a  (1,1) tensor field $J$ such that $J^{2}=-\mathrm{Id}$.
Specifically, we set 
\be\label{eqn: almost complex structure on statistical manifolds}
J\,=\,\alpha^{j}\,\otimes\,\Y_{j} - \beta^{j}\,\otimes\,\X_{j},
\ee
and a direct computation shows that $J^{2}\,=\,-\mathrm{Id}$.


Note that we do not require any integrability property for $J$, and this means that $J$ is not in general a proper complex structure on $M\times M$ like the tensor field $J_{\pi}$ on the complex projective space in section \ref{sec: pure states}, and in the literature it is called quasi-complex structure.
However, the procedure outlined at the end of section \ref{sec: pure states} relies only on the fact that $J^{2}\,=\,-\mathrm{Id}$ and on the compatibility between $J$ and $F$ expressed by equation \eqref{eqn: compatibility between F,J and omega}, and not on the integrability properties of $J$.
Therefore, if $F$ is any function on $M\times M$, we may still define a closed, antisymmetric 2-form $\omega^{J}_{F}$ on $M\times M$ setting
\be\label{eqn: symplectic-like form}
\begin{split}
\omega^{J}_{F}&\,:=\,\dd\,\dd_{J }F\,=\,\dd\,\left(J\,\circ \,\dd F\right)\,=\, \\
&\,=\, \mathrm{d}\,\left(\Y_{j}(F)\,\alpha^{j} - \X_{j}(F)\,\beta^{j}\right)\,=\, \\
&\,=\, \left(L_{\X_{j}}L_{\Y_{k}}(F)  - \frac{1}{2}\,\Y_{l}(F)\,\alpha^{l}([\X_{j},\X_{k}]) \right)\,\alpha^{j}\wedge\alpha^{k} + \\
& \,\;\;\; + \left( L_{\X_{k}}L_{\Y_{j}}(F) + \frac{1}{2}\,\X_{l}(F)\,\beta^{l}([\Y_{j},\Y_{k}]) \right)\,\beta^{j}\wedge\beta^{k} + \\
&\, \;\;\;\;+ \left(L_{\Y_{k}}L_{\Y_{j}}(F) + L_{\X_{j}}L_{\X_{k}}(F)\right)\,\beta^{k}\wedge \alpha^{j}  \,
\end{split}
\ee
and a (0,2) tensor field $g^{J}_{F}$ on $M\times M$ by setting
\be
g^{J}_{F}(Z,W)\,:=\,\omega^{J}_{F}(J(Z),W) 
\ee
for all vector fields $Z,W$ on $M\times M$. An explicit computation leads to the following expression  
\be\label{eqn: metric-like form}
\begin{split}
g^{J}_{F}&\,=\,\left(L_{\Y_{j}}L_{\Y_{k}}(F) + L_{\X_{k}}L_{\X_{j}}(F)\right)\,\left(\alpha^{j}\otimes \alpha^{k} + \beta^{k}\otimes\beta^{j}\right)+\\
& +\left( \X_{l}(F)\,\beta^{l}([\Y_{j},\Y_{k}])  -  L_{\X_{k}}L_{\Y_{j}}(F) + L_{\X_{j}}L_{\Y_{k}}(F)\right)\,\alpha^{j}\otimes\beta^{k}  + \\
&+ \left(\Y_{l}(F)\,\alpha^{l}([\X_{j},\X_{k}]) -   L_{\X_{j}}L_{\Y_{k}}(F) + L_{\X_{k}}L_{\Y_{j}}(F)\right)\,\beta^{j}\otimes\alpha^{k}  \\
\end{split}
\ee

This tensor will be symmetric whenever $J$ and $F$ satisfy the compatibility condition given in equation \eqref{eqn: compatibility between F,J and omega}.

The possibility of obtaining a pre-symplectic form on $M\times M$ has also been discussed in \cite{BN-J-1997,Noda-2011,Z-L-2013}.
The main difference between these works and our is that our considerations are  completely coordinate-independent and are   valid for all parallelizable statistical manifolds. Moreover, we want to stress the importance of the almost complex structure $J$, whose definition is coordinate independent, and its relation with the K\"{a}hler geometry of quantum mechanics as the basic building blocks of our procedure.
This is in line with the idea that the quantum setting, being ``more fundamental'', should cast its light on the classical setting inspiring the construction of new geometrical structures on the latter, and not vice-versa.


At this point, we may consider the diagonal immersion $i_{d} $   and take the pullback of $\omega^{J}_{F}$ and $g^{J}_{F}$ to $M$ to obtain  
\be\label{eqn: symplectic-like form on M}
\begin{split}
i_{d}^{*}\omega^{J}_{F}&\,=\,  \left.\left(L_{\X_{j}}L_{\Y_{k}}(F) - L_{\X_{k}}L_{\Y_{j}}(F) \right)\right|_{diag}\,\theta^{j}\wedge\theta^{k} + \\
& \,\;\;\; +  \left.\left(L_{\Y_{k}}L_{\Y_{j}}(F) + L_{\X_{j}}L_{\X_{k}}(F)\right)\right|_{diag}\,\theta^{j}\wedge \theta^{k} - \\
&\, \;\;\;\;- \frac{1}{2}\,\left.\left(\Y_{l}(F)\,\alpha^{l}([\X_{j},\X_{k}]) + \X_{l}(F)\,\beta^{l}([\Y_{j},\Y_{k}]) \right)\right|_{diag}\,\theta^{j}\wedge\theta^{k}  \,
\end{split}
\ee
and
\be
\begin{split}
i_{d}^{*}g^{J}_{F}&\,=\,  \left.\left(L_{\Y_{j}}L_{\Y_{k}}(F)  + L_{\X_{k}}L_{\X_{j}}(F)  \right)\right|_{diag}\, \theta^{j}\otimes_{s} \theta^{k}+ \\
& \,\;\;\;+\frac{1}{2}\left.\left( \Y_{l}(F)\,\alpha^{l}([\X_{j},\X_{k}]) + \X_{l}(F)\,\beta^{l}([\Y_{j},\Y_{k}]) \right)\right|_{diag}\,\theta^{j}\wedge \theta^{k} ,
\end{split}
\ee

where we used equation \eqref{eqn: important pullbacks}, and we set $ \theta^{j}\otimes_{s} \theta^{k} =  \theta^{j}\otimes \theta^{k} + \theta^{k}\otimes\theta^{j} $.

According to the results in section 2 of \cite{C-DC-L-M-M-V-V-2018}, $F$ is a divergence function if we have
\be
\left.\left(L_{\X_{j}}F\right)\right|_{diag}\,=\,\left.\left(L_{\Y_{j}}F\right)\right|_{diag}\,=\,0\;\;\forall j=1,...,\mathrm{dim}(M).
\ee
Then, $F$ is such that
\be
\begin{split}
\left.\left(L_{\X_{j}}L_{\X_{k}}F\right)\right|_{diag} &\,=\,\left.\left(L_{\X_{k}}L_{\X_{j}}F\right)\right|_{diag}\,=\,\left.\left(L_{\Y_{k}}L_{\Y_{j}}F\right)\right|_{diag}\,=\,\left.\left(L_{\Y_{j}}L_{\Y_{k}}F\right)\right|_{diag}\,=\,\\
&\,=\,-\left.\left(L_{\Y_{j}}L_{\X_{k}}F\right)\right|_{diag}\,=\,-\left.\left(L_{\X_{k}}L_{\Y_{j}}F\right)\right|_{diag}\,=\\
&=\,-\left.\left(L_{\X_{j}}L_{\Y_{k}}F\right)\right|_{diag}\,=\,-\left.\left(L_{\Y_{k}}L_{\X_{j}}F\right)\right|_{diag},
\end{split}
\ee
and we obtain 
\be\label{eqn: symplectic-like form on M if F is a potential function}
\begin{split}
i_{d}^{*}\omega^{J}_{F}&\,=\,   0,
\end{split}
\ee
and 
\be\label{eqn: metric-like form on M if F is a potential function}
\begin{split}
i_{d}^{*}g^{J}_{F}&\,=\,  2\left.\left(L_{\X_{j}}L_{\X_{k}}(F)  \right)\right|_{diag}\, \theta^{j}\otimes_{s} \theta^{k}\,=\, \\
&\,=\,  2\left.\left(L_{\Y_{j}}L_{\Y_{k}}(F)   \right)\right|_{diag}\, \theta^{j}\otimes_{s} \theta^{k}\,=\, \\
&\,=\, - 2\left.\left(L_{\X_{j}}L_{\Y_{k}}(F)  \right)\right|_{diag}\, \theta^{j}\otimes_{s} \theta^{k} ,
\end{split}
\ee
where we used proposition 5 in \cite{C-DC-L-M-M-V-V-2018}.
By comparing equation \eqref{eqn: metric-like form on M if F is a potential function} with the results of proposition 5 in \cite{C-DC-L-M-M-V-V-2018}, we obtain that the metric-like tensor on $M$ extracted from the function $F$ by means of the almost complex structure $J$ as explained in this section coincides with the metric-like tensor on $M$ extracted from the function $F$  by means of the procedure outlined in \cite{C-DC-L-M-M-V-V-2018}.
At this point, if $\mathfrak{J}$ is the almost complex structure associated with another choice of global bases on $M\times M$, a direct computation shows that we have
\be
\begin{split}
i_{d}^{*}\omega^{\mathfrak{J}}_{F}&\,=\,i_{d}^{*}\omega^{J}_{F} \,=\,0 \\
i_{d}^{*} g^{\mathfrak{J}}_{F}&\,=\,i_{d}^{*} g^{J}_{F}. 
\end{split}
\ee
where the last equality follows upon a direct comparison with the results in \cite{C-DC-L-M-M-V-V-2018}. 
Note that, in order for the previous equations to be true, we must impose that $F$ is a divergence function.


\vsp

We will now apply our considerations to the case of an N-simplex where the function $F$ is taken to be the Kullback-Leibler relative entropy 
\be
S_{KL}(\mathbf{p},\mathbf{q})\,=\,\sum_{j=1}^{N}\,p^{j}\ln\left(\frac{p^{j}}{q^{j}}\right) .
\ee
 
In the open interior of the simplex we have the basis $\{P_{j}\}_{j=1,...,(N-1)}$ of vector fields given by
\be
P_{j}\,=\,\frac{\partial}{\partial p^{j}} - \frac{\partial}{\partial p^{j+1}}, 
\ee
and its associated dual basis of 1-forms $\{\vartheta^{j}\}_{j=1,...,(N-1)}$ given by
\be
\vartheta^{j}\,=\,\frac{1}{2}\left(\mathrm{d}p^{j} - \mathrm{d}p^{j+1} - \sum_{k\neq j,j+1} \mathrm{d}p^{k}\right) .  
\ee

Denoting by $\{\p_{j},\Q_{j}\}_{j=1,...,(N-1)}$ and $\{\alpha^{j},\beta^{j}\}_{j=1,...,(N-1)}$ the basis of vector fields and one-forms on $\Delta_{+}\times\Delta_{+}$ obtained by $\{P_{j}\}_{j=1,...,(N-1)}$ and $\{\vartheta^{j}\}_{j=1,...,(N-1)}$, we first note that 
\be
[\p_{j},\p_{k}]=[\p_{j},\Q_{k}]=[\Q_{j},\Q_{k}]=0,
\ee
and then we directly compute
\be
\begin{split}
\omega^{J}_{KL}\,=\, \frac{1}{q^{l}} &\left(\delta_{kl}(\delta_{jl} - \delta_{(j+1)l}) - \delta_{(k+1)l}(\delta_{jl} - \delta_{(j+1)l})   \right)\,\cdot \\
\cdot & \left(\alpha^{j}\wedge\alpha^{k} + \beta^{j}\wedge\beta^{k} + 2\beta^{k}\wedge \alpha^{j}\right) \,, 
\end{split}
\ee
and
\be
\begin{split}
g^{J}_{F}&\,=\,\frac{2}{q^{l}}\left(\delta_{kl}(\delta_{jl} - \delta_{(j+1)l}) - \delta_{(k+1)l}(\delta_{jl} - \delta_{(j+1)l})   \right)\,\left(\alpha^{j}\otimes \alpha^{k} + \beta^{k}\otimes\beta^{j}\right),
\end{split}
\ee
so that on $\Delta_{+}$ we obtain
\be 
\begin{split}
i_{d}^{*}\omega^{J}_{KL}&\,=\,  0,
\end{split}
\ee
and 
\be
\begin{split}
i_{d}^{*}g^{J}_{KL}& \,=\,    2\,\left(\frac{\delta_{jk} - \delta_{(j+1)k}}{p^{k}}  + \frac{\delta_{(j+1)(k+1)} - \delta_{j(k+1)}}{p^{k+1}}\right) \,\vartheta^{j}\otimes_{s}\vartheta^{k} .
\end{split}
\ee
Concerning $i_{d}^{*}g^{J}_{KL}$, we immediately see that
\be
i_{d}^{*}g^{J}_{KL}(P_{j},P_{j})\,=\, 2\,\left(\frac{1}{p^{j}} + \frac{1}{p^{j+1}}\right),
\ee
while, if $k+1<j$ or $j+1<k$ we have 
\be
i_{d}^{*}g^{J}_{KL}(P_{j},P_{k})\,=\,  0,
\ee
and, if $k+1=j$ we have
\be
i_{d}^{*}g^{J}_{KL}(P_{j},P_{k})\,=\,  -\frac{2}{p^{j}},
\ee
and, if $j+1=k$ we have
\be
i_{d}^{*}g^{J}_{KL}(P_{j},P_{k})\,=\,  -\frac{2}{p^{j+1}}.
\ee
Then, a direct comparison using the explicit form of the $P_{j}$'s and the explicit form of the Fisher-Rao metric tensor $g_{FR}$ given in equation \eqref{eqn: F-R metric on the simplex} shows that
\be
i_{d}^{*}g^{J}_{KL} \,=\, 2\,g_{FR}\,.
\ee


\section{Information geometry of mixed states}\label{sec: mixed states}

In section \ref{sec: pure states}, we reviewed how classical probability distributions may be immersed into the complex projective space by replacing a probability vector with a suitable notion of probability amplitude.
Here, we want to develop a similar, but different, idea  in the context of mixed quantum states.
Pictorially speaking, we want to quantize classical probability distributions in order to obtain quantum states.
\begin{remark}
It is possible to pass from the concept of pure quantum ``states'' to the concept of quantum ``amplitudes''. In other words, a sort of operator-valued square root of a pure state can be introduced, if one considers the set of rank-one operators. Indeed, let $| \psi \rangle$ and $|\phi \rangle$ be normalized vectors in a Hilbert space $\mathcal{H}$, such that $\langle \psi | \phi \rangle \neq 0$ and $\rho_{\psi} = | \psi \rangle \langle \psi |$ and $\rho_{\phi}=|\phi \rangle \langle \phi |$ be the corresponding pure states in $\mathbb{P}(\mathcal{H})$. Let $\rho_{\psi \phi} = | \psi \rangle \langle \phi |$ be a rank one operator. Then, the following relations hold true:
\be
\begin{split}
\rho_{\psi \phi}\rho^*_{\psi \phi}& = \rho_{\psi} \\
 \rho^*_{\psi \phi}\rho_{\psi \phi}& = \rho_{\phi}\,,
\end{split}  
\ee
which amounts to say that the rank one operator $\rho_{\psi \phi}$ is a square root for the two states $\rho_{\psi}$ and $\rho_{\phi}$. It is worth noticing that this rank-one operator defines a transition amplitude because of the non-othogonality condition. These rules coincide with the algebraic structure underlying Schwinger's approach to quantum mechanics\cite{Schwinger-QM}, an approach which is based on the concept of selective measurements. It is possible to prove\cite{C-I-M-SchwingerI} that these basic elements satisfy the defining properties of a groupoid, each selective measurement describing a transition between outcomes of an experiment performed on a quantum system (for instance the outcomes of a Stern-Gerlach experiment on a beam of atoms). In this framework the rank-one operator $\rho_{\psi \phi}$ describes a selective measurement between the outcomes of two ``non-compatible'' experiments (for more details on this formulation see \cite{C-I-M-SchwingerI,C-I-M-SchwingerII,C-I-M-SchwingerIII}). 

A similar procedure may be implemented also for mixed states, see Section 2.3 in \cite{A-F-M-2019}. However, here we shall limit ourselves to mixed quantum states and do not consider their ``square root''.   
\end{remark}
In the quantum information theory of finite-level quantum systems, the role of probability distributions is played by the density operators on the Hilbert space $\hh$ of the system under investigation (see \cite{A-K-2019,BN-G-J-2003,B-Z-2006,Naudts-2018,Suzuki-2019}).
By fixing a basis $\{e_{j}\}$ in $\hh$, we may realize every density operator $\rho$ as a ``density matrix'' with respect to the chosen basis.
Then, we may consider a natural immersion of a probability vector $\mathbf{p}=(p_{1},...,p_{N})$ into the space of density matrices given by
\be\label{eqn; probability matrix}
\mathbf{p}\,\mapsto\,\rho_{p}\,=\,p^{j}\,E_{j}\,,
\ee
where $E_{j}=|e_{j}\rangle\langle e_{j}|$.
In this way, we realize a probability vector as a diagonal matrix with respect to the  basis  $\{e_{j}\}$ in $\hh$.
Here, since we shifted our attention from pure states to mixed states,   the analogue of the phase factor used in section \ref{sec: pure states} (i.e., and element of the unitary group $U(1)$) is an element $\mathbf{U}$ of the special  unitary group $\mathcal{SU}(\hh)$ of the Hilbert space of the system.
Therefore, by acting with this ``generalized phase factor'', we obtain the density operator
\be
\rho(\mathbf{U},\mathbf{p})\,:=\,\mathbf{U}\,\rho_{p}\,\mathbf{U}^{\dagger}\,
\ee
which is a density operator such that its vector of eigenvalues is in  one-to-one correspondence with $\mathbf{p}$ modulo an action of the permutation group.
In this way, we have just obtained a covering of the space of states (density operators) $\stsp$ of $\hh$ in terms of the space
\be
\widetilde{\mathcal{M}}\,=\,\mathcal{SU}(\hh)\,\times\,\Delta,
\ee
where $\Delta$ denotes the simplex of $N$-probability vectors, by means of the map $\widetilde{\pi}\colon\widetilde{\mathcal{M}}\lra\stsp$ given by
\be
\widetilde{\pi}(\mathbf{U},\mathbf{p})\,:=\,\rho(\mathbf{U},\mathbf{p})\,.
\ee 
By considering only probability vectors with $p_{j}>0$ for all $j=1,..,N$, we obtain a map $\pi$ from the smooth manifold
\be
\mathcal{M}\,=\,\mathcal{SU}(\hh)\,\times\,\Delta_{+},
\ee
where $\Delta_{+}$ denotes the open interior of the simplex of $N$-probability vectors, to the smooth manifold $\stsp_{+}$ of faithful states (invertible density operators) on $\hh$ given by the restriction of $\widetilde{\pi}$ to $\mathcal{M}$.
According to \cite{C-DC-L-M-M-V-V-2018}, the map $\pi$ is differentiable and it is a submersion at each point $(\mathbf{U},\mathbf{p})\in\mathcal{M}$ for which the vector $\mathbf{p}$ is such that $p_{j}\neq p_{k}$ for $j\neq k$.

The possibility of working on $\mathcal{M}$ turns out to be particularly useful when we need to perform explicit computations regarding geometrical structures on $\stsp_{+}$.
For instance, $\mathcal{M}$ is a parallelizable manifold, and thus, the theory developed in section \ref{sec: statistical manifolds} applies.
Therefore, if $S\colon\stsp_{+}\times\stsp_{+}\lra\mathbb{R}$ is a relative quantum entropy, e.g., the von Neumann-Umegaki relative entropy, we may consider its pullback $S_{\pi}$ to $\mathcal{M}\times\mathcal{M}$, and compute the geometrical tensors associated with it as explained in section \ref{sec: statistical manifolds}.
The symmetric tensor thus obtained will be the pullback on $\mathcal{M}\times\mathcal{M}$ of the symmetric tensor on $\stsp_{+}\times\stsp_{+}$ that we may obtain by applying one of the general procedures described in \cite{C-DC-L-M-M-V-V-2018,G-G-K-M-2019,L-M-M-V-V-2018}.
Furthermore,  the differential calculus available on $\mathcal{M}\times\mathcal{M}$ is considerably simple once we note that $\mathcal{M}$ is just the Cartesian product of a Lie group with an open set of an  affine space, and this makes some computations particularly explicit and clear.
Then, since the probability distributions are embedded in $\mathcal{M}$ in a manifest way, we get an easier comparison between geometrical structures on quantum states and geometrical structures on classical probabilities in the spirit of section \ref{sec: pure states} and of the work  \cite{F-K-M-M-S-V-2010}.

To better illustrate this point, we will now explicitely work out the details of the computation of the metric tensor on $\mathcal{M}$ for the case of the von Neumann-Umegaki relative entropy defined by
\be
S(\rho,\,\sigma)\,:=\,{\rm Tr }\left(\rho(\ln(\rho) - \ln(\sigma)\right)\,.
\ee
Note that, however, a similar procedure may be also considered for the family of Tsallis entropies (see \cite{M-M-V-V-2017}), and for the family of $(\alpha-z)$-Renyi relative entropies  \cite{C-DC-L-M-M-V-V-2018}.

For the sake of simplicity, we omit the explicit expression of the pre-symplectic form and the symmetric $(0,2)$ tensor on $\mathcal{M}\times\mathcal{M}$.
Indeed, the explicit expressions of these objects turn out to be particularly long, and the computations that we are about to perform in order to compute the metric tensor on $\mathcal{M}$ already give enough details to obtain the tensors on $\mathcal{M}\times\mathcal{M}$ by means of equations \eqref{eqn: symplectic-like form} and \eqref{eqn: metric-like form}.

\vsp

First of all, we need to introduce a global basis of vector fields on $\mathcal{M}$ and its dually-related basis of 1-forms.
At this purpose, we will exploit the product structure of $\mathcal{M}=\mathcal{SU}(\hh)\,\times\,\Delta_{+}$ to select the basis $\{X_{j};P_{k}\}$ of vector fields where the $X_{j}$'s are left-invariant vector fields ``on'' $\mathcal{SU}(\hh)$,  and the $P_{k}$'s are the vector fields on $\Delta_{+}$ introduced in the last part of section \ref{sec: statistical manifolds}.
Then, the dual basis will be written as $\{\theta^{j};\vartheta^{k}\}$, while the basis of vector fields on $\mathcal{M}\times\mathcal{M}$ is written as $\{\X_{j},\Y_{j};\p_{k},\Q_{k}\}$, and its dual basis of 1-forms as $\{\alpha^{j},\beta^{j};\zeta^{k},\eta^{k}\}$.

Setting $\rho_{0}\equiv\rho_{p}$ for $\mathbf{p}\in\Delta_{+}$ (see equation \eqref{eqn; probability matrix}), the pullback of $S$ to $\mathcal{M}$ can be written as
\be
S_{\pi}(\mathbf{U},\rho_{0};\mathbf{V},\sigma_{0})\,=\,{\rm Tr }\left(\rho_{0}\,\ln(\rho_{0})\right) - {\rm Tr }\left(\mathbf{U}\,\rho_{0}\,\mathbf{U}^{\dagger}\,\,\mathbf{V}\,\ln(\sigma_{0})\,\mathbf{V}^{\dagger}\right),
\ee
where we exploited the fact that $\ln(\mathbf{U}A\mathbf{U}^{\dagger})\,=\,\mathbf{U}\ln(A)\mathbf{U}^{\dagger}$ for every invertible self-adjoint operator $A$ because $\ln$ is an analytic function.

According to equation \eqref{eqn: metric-like form on M if F is a potential function}, we need to compute the quantities
\be
\left.\left(L_{\X_{k}}L_{\Y_{j}}(S_{\pi}) \right)\right|_{diag},\;\;\left.\left(L_{\X_{k}}L_{\Q_{j}}(S_{\pi}) \right)\right|_{diag},\;\;\left.\left(L_{\p_{k}}L_{\Q_{j}}(S_{\pi}) \right)\right|_{diag}
\ee
in order to obtain the tensor  $i_{d}^{*}g^{J}_{S}$.

In order to compute them, we first observe that the following matrix equality holds by direct computations
\be
\mathrm{d}\rho\,=\,\mathrm{d}\left(\mathbf{U}\,\rho_{0}\,\mathbf{U}^{\dagger}\right)\,=\,\mathbf{U}\,\left(\left[\mathbf{U}^{\dagger}\,\mathrm{d}\mathbf{U},\rho_{0}\right] + \mathrm{d}\rho_{0}\right)\,\mathbf{U}^{\dagger}, 
\ee
where $\mathbf{U}^{\dagger}\,\mathrm{d}\mathbf{U}$ is the left-invariant, Maurer-Cartan form on $\mathcal{SU}(\hh)$.
Note that, If we fix an orthonormal  basis $\{\tau_{j}\}$ of matrices in the Lie algebra of $\mathcal{SU}(\hh)$ (w.r.t the Cartan-Killing form), the Maurer-Cartan form can be written as
\be
\mathbf{U}^{\dagger}\,\mathrm{d}\mathbf{U}\,=\,\theta^{j}\,\tau_{j}.
\ee
Furthermore, setting $E_{j}=|e_{j}\rangle\langle e_{j}|$  where $\{e_{j}\}$ is the orthonormal basis in $\hh$ in terms of which the unfolding map $\pi$ is written, we have
\be
\mathrm{d}\rho_{0}\,=\,\mathrm{d}p^{j}\,E_{j}\,,
\ee
so that\footnote{Recall that, for the basis on the $N$-simplex, the indexes run from $1$ to $N-1$.}
\be
\mathrm{d}\rho_{0}(P_{k})\,=\,E_{j} - E_{j+1}\,,
\ee 
In the following, we will select a basis $\{\tau_{k}\}_{1,...,(N^{2}-1)}$ in the Lie algebra of $\mathcal{SU}(\hh)$ in such a way that the $\imath E_{j}$'s are part of this algebra, and thus form a Cartan subalgebra.

A direct computation shows that
\be
\begin{split}
L_{\X_{k}}L_{\Y_{j}}(S_{\pi})&\,=\,-L_{\X_{k}}\left( {\rm Tr }\left(\mathbf{U}\,\rho_{0}\,\mathbf{U}^{\dagger}\,\mathbf{V}\,\left[(\mathbf{V}^{\dagger}\mathrm{d}\mathbf{V})(\Y_{j}),\, \ln(\sigma_{0})\right] \,\mathbf{V}^{\dagger}\right)\right)\,=\, \\
&\,=\,-L_{\X_{k}}\left( {\rm Tr }\left(\mathbf{U}\,\rho_{0}\,\mathbf{U}^{\dagger}\,\,\mathbf{V}\,\left[\tau_{j},\, \ln(\sigma_{0}) \right]\,\,\mathbf{V}^{\dagger}\right)\right)\,=\,\\
&\,=\, -  {\rm Tr }\left(\,\mathbf{U}\,\left[(\mathbf{U}^{\dagger}\mathrm{d}\mathbf{U})(\X_{k}),\, \rho_{0}\right] \,\mathbf{U}^{\dagger}\,\,\mathbf{V}\,\left[\tau_{j},\, \ln(\sigma_{0}) \right]\,\,\mathbf{V}^{\dagger}\right) \,=\, \\
&\,=\, -  {\rm Tr }\left(\mathbf{U}\,\left[\tau_{k},\, \rho_{0}\right] \,\mathbf{U}^{\dagger}\,\,\mathbf{V}\,\left[\tau_{j},\, \ln(\sigma_{0}) \right]\,\,\mathbf{V}^{\dagger}\right) ,
\end{split}
\ee
so that it is
\be
\left.\left(L_{\X_{k}}L_{\Y_{j}}(S_{\pi})\right)\right|_{diag}\,=\,{\rm Tr }\left(\left[\rho_{0},\,\tau_{k}\right]\,\left[\tau_{j},\,\ln(\rho_{0}) \right]\right) \,.
\ee

Similarly, we have
\be
\begin{split}
L_{\X_{k}}L_{\Q_{j}}(S_{\pi})&\,=\,-L_{\X_{k}}\left( {\rm Tr }\left(\mathbf{U}\,\rho_{0}\,\mathbf{U}^{\dagger}\,\,\mathbf{V}\,\left(\mathrm{d}(\ln(\sigma_{0}))\right)(\Q_{j})\,\mathbf{V}^{\dagger}\right)\right)\,=\, \\
&\,=\,-  {\rm Tr }\left(\mathbf{U}\,\left[\tau_{k},\, \rho_{0}\right] \,\mathbf{U}^{\dagger}\,\,\mathbf{V}\,\left(\mathrm{d}(\ln(\sigma_{0}))\right)(\Q_{j})\,\mathbf{V}^{\dagger}\right)  \,=\, \\
&\,=\, -  {\rm Tr }\left(\mathbf{U}\,\left[\tau_{k},\, \rho_{0}\right] \,\mathbf{U}^{\dagger}\,\,\,\mathbf{V}\, \sigma_{0}^{-1}\,\left(E_{j} - E_{j+1}\right) \,\mathbf{V}^{\dagger}\right),
\end{split}
\ee
so that it is
\be
\left.\left(L_{\X_{k}}L_{\Q_{j}}(S_{\pi})\right)\right|_{diag}\,=\, {\rm Tr }\left(\left[\rho_{0},\,\tau_{k}\right]\,\rho_{0}^{-1}\,\left(E_{j} - E_{j+1}\right)\right)\,=\,0\,,
\ee
and we have
\be
\begin{split}
L_{\p_{k}}L_{\Q_{j}}(S_{\pi})&\,=\,-L_{\Q_{k}}\left( {\rm Tr }\left(\mathbf{U}\,\rho_{0}\,\mathbf{U}^{\dagger}\,\,\mathbf{V}\,\left(\mathrm{d}(\ln(\sigma_{0}))\right)(\Q_{j})\,\mathbf{V}^{\dagger}\right)\right)\,=\, \\
&\,=\,-L_{\Q_{k}}\left(  {\rm Tr }\left(\mathbf{U}\,\rho_{0}\,\mathbf{U}^{\dagger}\,\, \mathbf{V}\,\sigma_{0}^{-1}\,\left(E_{j} - E_{j+1}\right)\,\mathbf{V}^{\dagger}\right)\right) \,=\, \\
&\,=\,-   {\rm Tr }\left(\mathbf{U}\,\left(E_{k} - E_{k+1}\right)\,\mathbf{U}^{\dagger}\,\,\mathbf{V}\,\sigma_{0}^{-1}\,\left(E_{j} - E_{j+1}\right)\,\mathbf{V}^{\dagger}\right)  ,
\end{split}
\ee
so that it is
\be
\left.\left(L_{\p_{k}}L_{\Q_{j}}(S_{\pi})\right)\right|_{diag}\,=\,-{\rm Tr }\left( \rho_{0}^{-1}\,\left(E_{k} - E_{k+1}\right)\,\left(E_{j} - E_{j+1}\right)\right)\,.
\ee

Collecting the results,  from equation \eqref{eqn: metric-like form on M if F is a potential function} we get
\be
\begin{split}
i_{d}^{*}g^{J}_{S}\,=\,&2\,{\rm Tr }\left( \rho_{0}^{-1}\,\left(E_{k} - E_{k+1}\right)\,\left(E_{j} - E_{j+1}\right)\right)\, \vartheta^{j}\otimes_{s}\vartheta^{k} +\\ 
+ &2\,{\rm Tr }\left(\left[\tau_{k},\,\rho_{0}\right]\,\left[\tau_{j},\,\ln(\rho_{0}) \right]\right) \,\theta^{j}\otimes_{s}\theta^{k}.
\end{split}
\ee
If we write 
\be
\rho_{0}\,=\,\sum_{j=1}^{N}\,p^{j}\,E_{j},
\ee
a direct comparison with the expression of the Fisher-Rao metric tensor $g_{FR}$ given at the end of section \ref{sec: statistical manifolds} shows that 
\be
g_{FR}\,=\,{\rm Tr }\left( \rho_{0}^{-1}\,\left(E_{k} - E_{k+1}\right)\,\left(E_{j} - E_{j+1}\right)\right)\, \vartheta^{j}\otimes_{s}\vartheta^{k}\,,
\ee
which means that the metric-like tensor on $\mathcal{M}$ reduces to (a multiple of) the Fisher-Rao metric tensor whenever we restrict to the ``classical part of the system'', i.e., to pairwise commuting matrices.

Now, we will show that there exists a basis in the Lie algebra of $\mathcal{SU}(\hh)$ for which the ``quantum part'' of $i_{d}^{*}g^{J}_{S}$ is diagonal.
Indeed, setting $\mathbf{E}_{jk}=|e_{j}\rangle\langle e_{k}|$, we obtain a basis $\{\mathbf{E}_{jk}\}_{j,k=1,...,N}$ of $\bh$.
Therefore, we can write
\be
\begin{split}
\rho_{0}&\,=\,\sum_{r} p^{r}\,\mathbf{E}_{rr}\,=\,\sum_{r}\,p^{r}\,E_{r}, \\
\tau_{j}&\,=\,\sum_{r,s}T^{rs}_{j}\,\mathbf{E}_{rs}\;\mbox{ with }\;T^{rs}_{j}\,=\,-\overline{T^{sr}_{j}},
\end{split}
\ee
so that
\be
\begin{split}
\left[\rho_{0},\tau_{k}\right]&\,=\,\sum_{r,s}\,\left(p^{s}-p^{r}\right)\,T^{rs}_{k}\,\mathbf{E}_{rs} \\
\left[\tau_{j},\ln(\rho_{0})\right]&\,=\,\sum_{a,b}\,\ln\left(\frac{p^{a}}{p^{b}}\right)\,T^{ab}_{j}\,\mathbf{E}_{ab},
\end{split}
\ee
and thus
\be\label{eqn: quantum part of the metric of vN-U relative entropy}
\begin{split}
{\rm Tr }\left(\left[\tau_{k},\,\rho_{0}\right]\,\left[\tau_{j},\,\ln(\rho_{0}) \right]\right)&\,=\,\sum_{r,s}\,\ln\left(\frac{p^{s}}{p^{r}}\right)\,\,\left(p^{r}-p^{s}\right)\,T^{rs}_{k}\,T^{sr}_{j}\,=\, \\
&\,=\,\sum_{r,s}\,\ln\left(\frac{p^{s}}{p^{r}}\right)\,\,\left(p^{r}-p^{s}\right)\,\frac{1}{2}\,\left(T^{rs}_{k}\,T^{sr}_{j} + T^{sr}_{k}\,T^{rs}_{j}\right)\,=\, \\
&\,=\,\sum_{r,s}\,\ln\left(\frac{p^{s}}{p^{r}}\right)\,\,\left(p^{r}-p^{s}\right)\,\frac{1}{2}\,\left(T^{rs}_{k}\,T^{sr}_{j} + \overline{T^{rs}_{k}\,T^{sr}_{j}}\right)\,=\, \\
&\,=\,\sum_{r,s}\,\ln\left(\frac{p^{s}}{p^{r}}\right)\,\,\left(p^{r}-p^{s}\right)\,\Re\,\left(T^{rs}_{k}\,T^{sr}_{j}\right) .
\end{split}
\ee
Now, we may chose a particular basis  in the Lie algebra of $\mathcal{SU}(\hh)$ which is splitted in three parts, namely, the elements of the basis are of three kinds, first, there are elements of the type
\be
\lambda_{j\mu}\,=\,\frac{\imath}{\sqrt{2}}\,\left(|j\rangle\langle \mu| + |\mu\rangle\langle j |\right)\,
\ee
where we always relabel the indexes in such a way that Greek ones are greater than the Latin ones, then, there are elements of the type
\be
\sigma_{j\mu}\,=\,-\frac{1}{\sqrt{2}}\,\left(|j\rangle\langle \mu| - |\mu\rangle\langle j |\right)
\ee
where we always relabel the indexes in such a way that Greek ones are greater than the Latin ones, and finally, there are elements of the type
\be
 \mathbf{e}_{j}\,=\, \alpha_{j}\,\left( j|j+1\rangle\langle j+1| - \sum_{r=1}^{j}\,|r\rangle\langle r|\right)
\ee
with $\alpha_{j}=\frac{1}{\sqrt{j(j+1)}}$.
Note that this basis is orthonormal with respect to the Cartan-Killing form on $\mathcal{SU}(\hh)$.
At this point, we write
\be\label{sigmatau}
\lambda_{j\mu}\,=\,\sum_{\alpha,\beta}\,M_{j\mu}^{\alpha\beta}\mathbf{E}_{\alpha\beta}\,,\;\;\;\; \sigma_{j\mu}\,=\,\sum_{\alpha,\beta}\,N_{j\mu}^{\alpha\beta}\mathbf{E}_{\alpha\beta}\,,\;\;\;\; \mathbf{e}_{\mu}\,=\,\sum_{\alpha,\beta}\,O_{\mu}^{\alpha\beta}\mathbf{E}_{\alpha\beta}\,,
\ee
from which it immediately follows that the quantity in equation \eqref{eqn: quantum part of the metric of vN-U relative entropy} is different from zero if and only if $\tau_{j}=\tau_{k}=\lambda_{l\mu}$ for some couple $(l,\mu)$, or  $\tau_{j}=\tau_{k}=\sigma_{l\mu}$ for some couple $(l,\mu)$.
From this, we conclude that, with respect to the basis of left-invariant vector fields and 1-forms associated with the basis $\{\lambda_{j\mu},\sigma_{j\mu},\mathbf{e}_{j}\}$ in the Lie algebra of $\mathcal{SU}(\hh)$,  the ``quantum part'' of $i_{d}^{*}g^{J}_{S}$ is diagonal.

Quite interestingly, this is also what happens when we consider the family of $\alpha$-z-Renyi-Relative-Entropies introduced in \cite{A-D-2015}.
Indeed, as it is shown in \cite{C-DC-L-M-M-V-V-2018}, the metric tensor is again diagonal with respect to  the basis of left-invariant vector fields and 1-forms associated with the basis $\{\lambda_{j\mu},\sigma_{j\mu},\mathbf{e}_{j}\}$ in the Lie algebra of $\mathcal{SU}(\hh)$. Furthermore, in the same paper it is shown also that the metric tensor derived from Von Neumann-Umegaki relative entropy is monotone with respect to quantum stochastic maps (completely positive and trace preserving maps on the $C^*$-algebra associated to the quantum system). According to Petz theorem, monotone metric tensor can be decomposed into the sum of two terms, a classical part which is the Fisher-Rao metric tensor, and a quantum term which is coupled to the classical one via a monotone function $f$ (a relation between this monotone function and the tomographic procedure to reconstruct a quantum state from the knowledge of different associated probability densities, has been proposed in \cite{L-M-M-V-V-2018,M-M-V-V-2017}). As already pointed out, Eq. \eqref{eqn: metric-like form on M if F is a potential function} shows that the metric tensor on $M$ obtained from a divergence function via a complex structure $J$ is proportional to the one obtained according to the procedure outlined in \cite{C-DC-L-M-M-V-V-2018}. Therefore, also the metric tensor that we got in this section satisfies the monotonicity property with respect to quantum stochastic maps.


\section{Concluding remarks}

Inspired by the geometry of pure quantum states, in this work, we presented the construction of a quasi-complex structure $J$ on the Cartesian product $M\times M$ of a parallelizable statistical manifold $M$.
By exploiting the geometrical properties of $J$, we defined a coordinate-free, algorithmic procedure to extract a symmetric, covariant (0,2) tensor $g_{F}$ and a presymplectic structure $\omega_{F}$ on $M\times M$ starting from a divergence function $F$ on $M\times M$.
In particular, in section \ref{sec: statistical manifolds}, we computed $\omega_{F}$ and  $g_F$ when $M$ is the open interior $\Delta_{+}$ of the $n$-simplex $\Delta$, and $F$ is the Kullback-Leibler divergence, and we proved that the pullback to $\Delta_{+}$ of the symmetric tensor field $g_{F}$ is a constant multiple of the Fisher-Rao metric tensor.
Then, in section \ref{sec: mixed states}, we considered the case in which $M$ is a covering of the manifold of faithful quantum states in finite-dimensions, that is, $M=\mathcal{SU}(\hh)\times\Delta_{+}$ where $\mathcal{SU}(\hh)$ is the special unitary group of the $n$-dimensional Hilbert space $\hh$ of the quantum system at hand, and $F$ is the von Neumann-Umegaki relative entropy.
In this case,  the metric tensor one obtains on $M=\mathcal{SU}(\hh)\times\Delta_{+}$ is splitted in two parts, one which ``lives'' on the classical part $\Delta_{+}$ and is a constant multiple of the Fisher-Rao metric tensor, and one which ``lives'' on the quantum part $\mathcal{SU}(\hh)$ and is a constant multiple of the symmetric covariant tensor extracted from the von Neumann-Umegaki relative entropy as done, for instance, in \cite{M-M-V-V-2017,C-DC-L-M-M-V-V-2018}.


\section{Acknowledgements}

F.D.C. would like to thank partial support provided by the MINECO research project MTM2017-84098-P and QUITEMAD++, S2018/TCS-A4342. G.M. acknowledges financial support from the Spanish Ministry of Economy and Competitiveness, through the Severo Ochoa Programme for Centres of Excellence in RD(SEV-2015/0554). G.M. would like to thank the support provided by the Santander/UC3M Excellence Chair Programme 2019/2020, and he is also a member of the Gruppo Nazionale di Fisica Matematica (INDAM),Italy.



\end{document}